# Effects of Electric Field Gradient on Sub-nanometer Spatial Resolution of Tip-enhanced Raman Spectroscopy

Lingyan Meng,[1,2] Zhilin Yang,[2,*] Jianing Chen,[1,3,*] Mengtao Sun[1,*]

[1]*Beijing National Laboratory for Condensed Matter Physics, Institute of Physics, Chinese Academy of Sciences, Beijing, 100190, China*

[2]*Department of Physics, Xiamen University, Xiamen, 361005, China*

[3] *Collaborative Innovation Center of Quantum Matter, Beijing, 100190, China*

Tip-enhanced Raman spectroscopy (TERS) with sub-nanometer spatial resolution has been recently demonstrated experimentally. However, the physical mechanism underlying is still under discussion. Here, we theoretically investigate the electric field gradient of a coupled tip-substrate system. Our calculations suggest that the ultra-high spatial resolution of TERS can be partially attributed to the electric field gradient effect owning to its tighter spatial confinement and sensitivity to the infrared (IR)-active of molecules. Particularly, in the case of TERS of flat-lying $H_2TBPP$ molecules，we find the electric field gradient enhancement is the dominating factor for the high spatial resolution, which qualitatively coincides with previous experimental report. Our theoretical study offers a new paradigm for understanding the mechanisms of the ultra-high spatial resolution demonstrated in tip-enhanced spectroscopy which is of importance but neglected.
.

Tip-enhanced Raman spectroscopy (TERS) is a particularly important spectral technique for molecule analysis on the nanometer scale because of its high detection sensitivity and spatial resolution [1-6]. Further improving of its spatial resolution is an important step to obtain highly resolved optical images of nanometric objects [7-11]. In TERS the near-field electromagnetic coupling of diffraction limited incident light to a metal or metalized scanning probe microscopy (SPM) tip leads to a huge localized electromagnetic field enhancement in the narrow nanogap between tip and substrate [12, 13]. The localized field enhancement yields an ultra-high spatial resolution breaking down the diffraction limit of light. The highest theoretically calculated spatial resolution of TERS can be 2 nm, which is closely related to the spatial extent of the localized electromagnetic field [14]. Higher order nonlinear effects are also considered to be related to such ultra-high resolution [15]. However, these physical mechanisms are not sufficient to comprehensively interpret the sub-nanometer spatial resolution in TERS. Researchers spared no effort to improve the TERS resolution experimentally [16-18], and recently, a spatial resolution of 0.5 nm for TERS has been demonstrated experimentally under ultrahigh vacuum and low temperature [1]. Although there are several theories proposed to understand this high spectral spatial resolution, the physical mechanism is ongoing under discussion.

Here we show by theoretical calculations that for flat-lying $H_2TBPP$ molecules and $H_2TBPP$ molecules with a small tilted angle, the electric field gradient has a larger effect than the electric field in terms of the spectral spatial resolution. The calculated spectral spatial resolution, Raman and infrared spectra coincide with the experimental results indicating the strong influence of electric field gradient on TERS [1].

In tip-enhanced Raman spectroscopy, both Raman-active modes and infrared (IR)-active modes can be simultaneously observed *in situ* [12, 19, 20]. The IR-active modes are attributed to the electric field gradient effect. Particularly, for a molecule placed in an inhomogeneous electromagnetic field, the Hamiltonian for the Raman spectra can be written as [21],

$$H = H_0 + H_1 + H_2 + H_3 + \cdots$$
$$= \alpha_{\alpha\beta} E_\beta E_\alpha + \frac{1}{3}\sum_{\alpha\beta\gamma} A_{\alpha,\beta\gamma} \nabla E_{\beta\gamma} E_\alpha + \frac{1}{3}\sum_{\alpha\beta\gamma} A_{\gamma,\alpha\beta} \nabla E_{\alpha\beta} E_\gamma + \frac{1}{3}\sum_{\alpha\beta\gamma\delta} C_{\alpha\beta,\gamma\delta} \nabla E_{\gamma\delta} \nabla E_{\alpha\beta} + \cdots \quad (1)$$

where $\alpha_{\alpha\beta}$, $A_{\alpha,\beta\gamma}$ and $C_{\alpha\beta,\gamma\delta}$ are the electric dipole-dipole polarizability, dipole-quadrupole polarizability and quadrupole-quadrupole polarizability, respectively, describing the distortion of the molecule by an external electric field. $E_\alpha$, $E_\beta$ and $E_\gamma$ are the external electric field. $\nabla E_{\alpha\beta}$, $\nabla E_{\gamma\delta}$ and $\nabla E_{\beta\gamma}$ are the electric field gradient. The first term of Eq. (1) accounts for the dipole Raman, and the higher order including $A_{\alpha,\beta\gamma}$ and $C_{\alpha\beta,\gamma\delta}$ account for the electric field gradient multiple Raman corresponding to IR-active modes [19,20, 22, 23]. The selection rules for the four terms in Eq. (1) can be obtained by the same manner in the following form [22],

$$\Gamma \in \Gamma_\mu = \Gamma_\theta \tag{2}$$

where $\Gamma$ is the irreducible representation.

According to Eq. (1), the intensities of vibrational modes of a molecule can be written as

$$\begin{aligned} I &= I_1 + I_2 + \cdots \\ &= (H_0 + H_1 + H_2 + H_3 + \cdots)^2 \\ &= \langle j|H_0|i\rangle^2 + 4\langle j|H_0|i\rangle\langle j|H_1|i\rangle + \cdots \end{aligned} \tag{3}$$

where the Raman shift is ignored, and $H_1 = H_2$ is omitted. Only the first two terms are big enough to be taken into account even in the high electric field gradient regions. Therefore, the intensity ratio of Raman-active mode to IR-active modes can be written as

$$\begin{aligned} \frac{I_1}{I_2} &= \frac{\langle j|H_0|i\rangle^2}{4\langle j|H_0|i\rangle\langle j|H_1|i\rangle} \\ &= \frac{(\alpha_{\alpha\beta} E_\beta E_\alpha)^2}{4 \times \alpha_{\alpha\beta} E_\beta E_\alpha \times \frac{1}{3}\sum_{\alpha\beta\gamma} A_{\alpha,\beta\gamma} \nabla E_{\beta\gamma} E_\alpha} \\ &= \frac{1}{4} \times \frac{E_\beta}{\nabla E_{\beta\gamma}} \times \left(\frac{1}{3}\sum_{\alpha\beta\gamma} \frac{A_{\alpha,\beta\gamma}}{\alpha_{\alpha\beta}}\right)^{-1} \end{aligned} \tag{4}$$

In Eq. (4), the contribution of the term $\frac{A_{\alpha,\beta\gamma}}{\alpha_{\alpha\beta}}$ is related to molecular polarizability, and the term $\frac{E_\beta}{\nabla E_{\beta\gamma}}$ represents the plasmonic contribution from the tip-substrate system. If we only consider the plasmon terms, then we obtained

$$\frac{I_1}{I_2} \propto \frac{1}{4} \times \frac{E_\beta}{\nabla E_{\beta\gamma}}$$

which describes the ratio of electric field enhancement to the electric field gradient enhancement.

To study in detail the electric field enhancement and the electric gradient enhancement in

TERS, we perform a numerical simulation, where a model of a conical gold tip on a silver substrate is employed. These calculations are carried out by the finite element method (FEM), which numerically solves the Maxwell's equations [24].

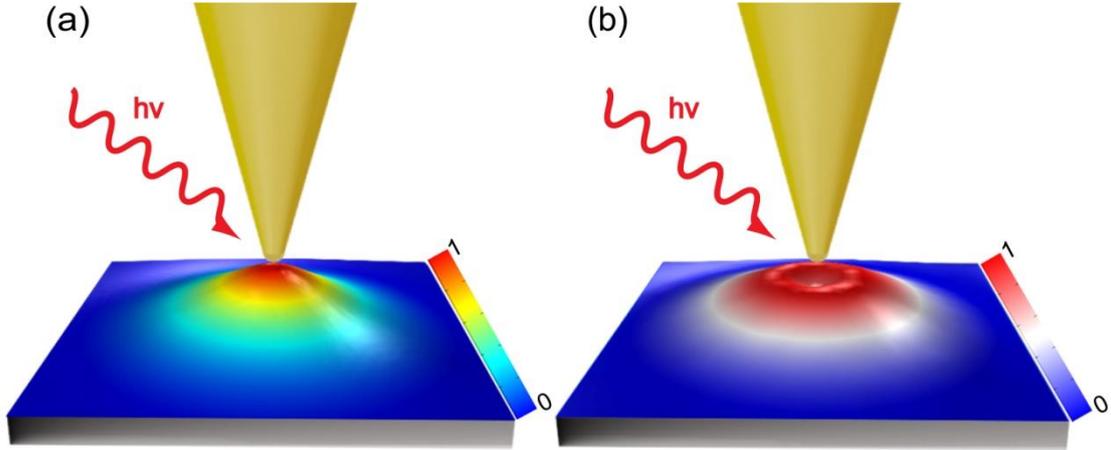

FIG. 1. (color online) Schematics of electric field (a) and electric field gradient (b) intensity distribution of the plane between the tip and substrate in TERS configuration.

The exact TERS configuration is shown in Fig.1 where a gold tip with final radius of 2 nm and a full cone angle $\phi$ is placed 1 nm above a silver substrate. The optical constants for Au and Ag were taken from Ref.[25]. The tip-substrate is illuminated with a p-polarized plane wave at an angle of $60^0$ relative to the tip, and its electric field amplitude is set at 1.0 V/m. Fig. 1(a) is the calculated intensity distribution of electric field and Fig.1(b) is counter part of the electric field gradient where the full cone angle is $20^o$. We observe different spatial pattern for the two quantities. The maximum of the electric field enhancement locates at the center directly below the tip, in contrast the maximum of the electric field gradient forms a ring shape below the tip. Generally, the Raman active modes arise from the contribution of electric field while the infrared active modes are attributed to the electric field gradient effect [12, 20]. These two fields indicate different active area for the dipole Raman and field gradient Raman.

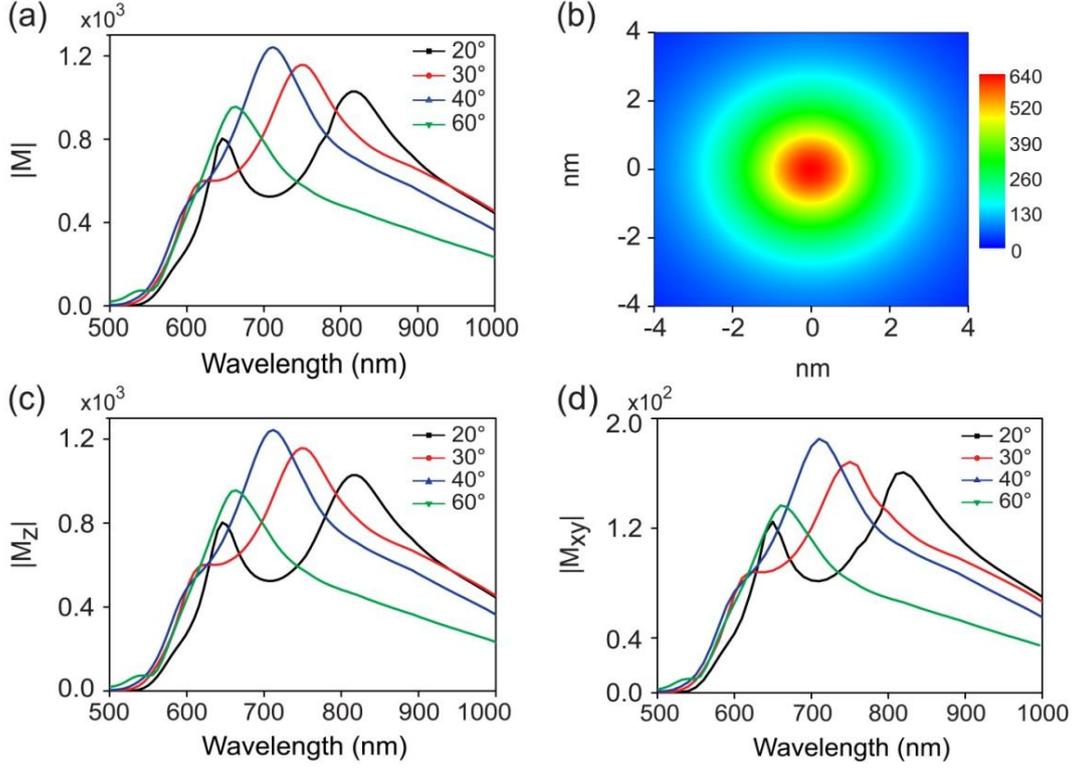

FIG. 2. (a) The maximum total electric field enhancement (defined as $|M| = |E_{loc}/E_{in}|$) as a function of wavelength and cone angle of the tip. (b) The total electric field distribution of the plane between the tip and substrate, where $\phi=20°$ and the incident wavelength is 632.8 nm. (c) The maximum vertical electric field enhancement ($|M_z|$), and (d) the maximum horizontal electric field enhancement ($|M_{xy}|$) as a function of wavelength and cone angle of the tip.

Fig. 2(a) shows the FEM simulation of the maximum total electric field enhancement (defined as the ratio between the maximum local field $E_{loc}$ and the incident field $E_{in}$ amplitude, $|M| = |E_{loc}/E_{in}|$) at the center of the nanogap between the tip and the substrate as a function of incident wavelength and cone angle of the tip. In all cases we observe two resonance behaviors which are attributed to the vertical (long wavelength) and the horizontal (short wavelength) dipole resonance respectively (see Fig. S1). Both the vertical and horizontal dipole resonances are red shift as decreasing the cone angle but with different speed. At $\phi=20°$, these two resonance are well separated exhibiting two discrete peaks.

We choose this tip ($\phi=20°$) and horizontal dipole resonance, which coincides with the vibration orientation of flat-lying $H_2TBPP$ molecules, as our research object to reveal the mechanism under the ultra-high spatial resolution of TERS. Fig. 2 (b) shows the total electric field distribution of the plane at the center of the tip and substrate with excitation wavelength of 632.8 nm. The figure clearly shows that the highest electric field enhancement is located at the center of the plane. It is believed that the spatial resolution of TERS is limited by the confinement of the electromagnetic field. Fig. 2(b) shows that this spatial extent is about 2 nm in diameter. Note that the TERS enhancement factor is proportional to the fourth power of the local electric field enhancement ($|E_{loc}/E_{in}|^4$), which reveals an even higher spatial resolution of TERS. In Fig. 2 (c) and 2 (d), we show the dependence of the maximum vertical and horizontal electric field enhancement, ($|M_z|$), and ($|M_{xy}|$) respectively, on the wavelength and cone angle of the tip. The plasmon resonances properties for both orientations are similar to that of the total electric field.

In Fig. 3 we study the electric field and its gradient field distributions, and their contributions to the TERS resolution in the horizontal orientation at the wavelength 632.8 nm. As shown in Fig. 3 (a), the highest horizontal electric field enhancement ($|M_{xy}|$) forms a ring shape with 0.7<r<1.7 nm (r is the distance away from the center of the plane) which has a maximum value around $1 \times 10^2$. In Fig.3 (b), the strongest electric field gradient also forms a ring shape with 0.7<r<1.7 nm showing the maximum of electric field gradient enhancement ($|\nabla M_{xy}|$) of 14.2 in atomic units. Fig. 3 (c) shows the distribution of the ratio of electric field enhancement to electric field gradient enhancement ($|M_{xy}/\nabla M_{xy}|$). The spatial pattern forms a wider ring shape showing the maximum ratio of 7.4. This value arises from the consideration of only the second term in Equation (1). If we also consider the third term which is equal to the second term, the ratio is 3.7.

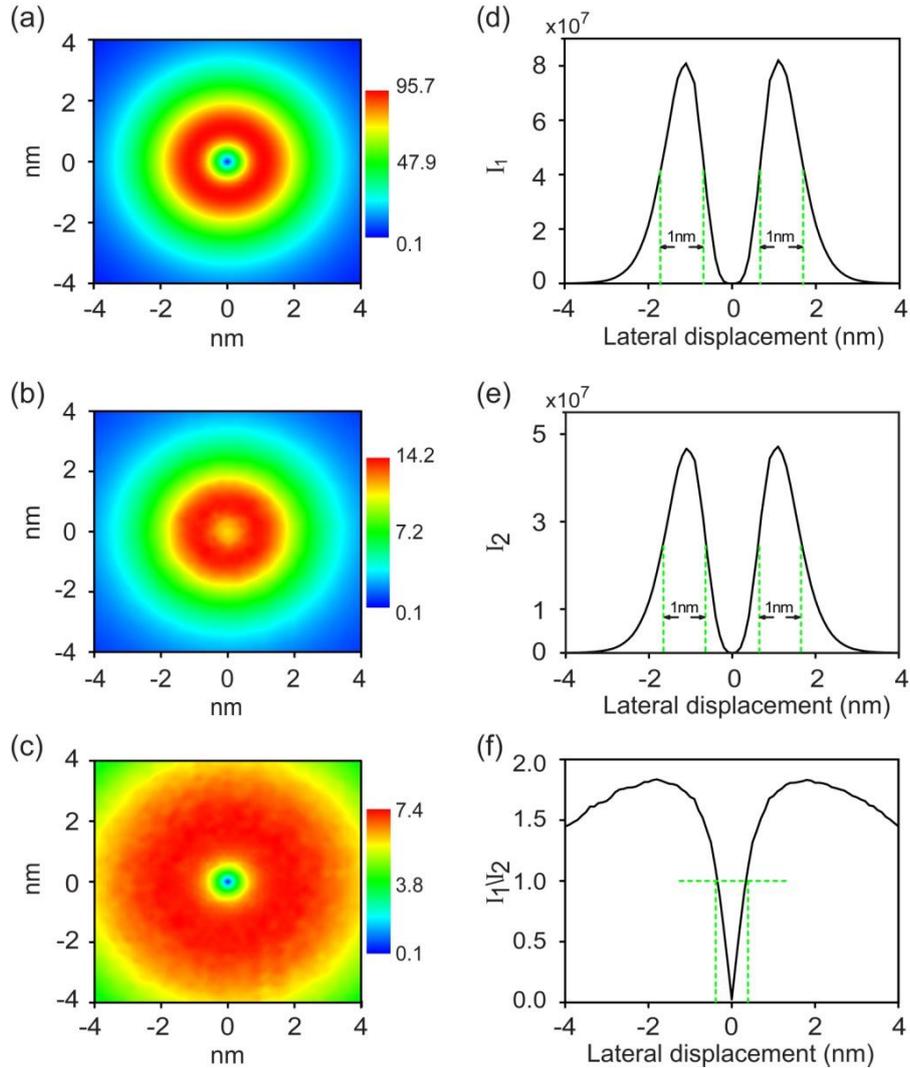

FIG.3. (color online) (a) The horizontal electric field and (b) horizontal electric field gradient distribution of the plane between the tip and substrate, where d=1 nm, $\phi=20°$. (c) The ratio of (a) over (b). (d) and (e) The $I_1$, $I_2$ and $I_1/I_2$ is plotted as a function of the lateral displacement which can reveal the spatial resolution of $I_1$, $I_2$ by full width at half maximum (FWHM). (f) $I_1/I_2$ is plotted as a function of the lateral displacement which describes the ratio of electric field to its gradient. The electric gradient is in unit of au. The excitation wavelength is 632.8 nm.

Fig. 3 (d)-(f) shows profiles of $I_1$, $I_2$ and $I_1/I_2$ plotted as function of the lateral displacement of the midway on the 2D plane. The spatial resolution is defined as the full width at half maximum (FWHM) in these profiles. As shown in Fig. 3d, the maximum $I_1$ is $8.0 \times 10^7$ and the spatial resolution can be as high as 1 nm. Fig. 3e shows that the maximum $I_2$ is $4.7 \times 10^7$ and the spatial resolution is 1 nm. In Fig. 3f, the maximum ratio of $I_1$ over $I_2$ is 1.8 located at $r = \pm 1.8$ nm. Importantly, the ratio $I_1/I_2$ is less than 1 at -0.4<r<0.4 nm, which reveals a more contribution of electric field gradient to the ultra-high spatial resolution of TERS than the electric field in the spatial extent of 0.8 nm in diameter directly below the tip, even though the same TERS resolutions of 1 nm are obtained from the two fields. If considering the combined effect of these two plasmon fields, we should obtain an even higher spatial resolution below 1 nm in TERS experiments.

It has been long believed that the strongest Raman signals of molecules can be achieved when the polarized direction of the external excitation electric field is in line with the vibration direction of the molecules. However, as these results demonstrate, the horizontal electric field gradient could have a more dominant contribution to the TERS resolution than the electric field for flat-lying $H_2$TBPP molecules.

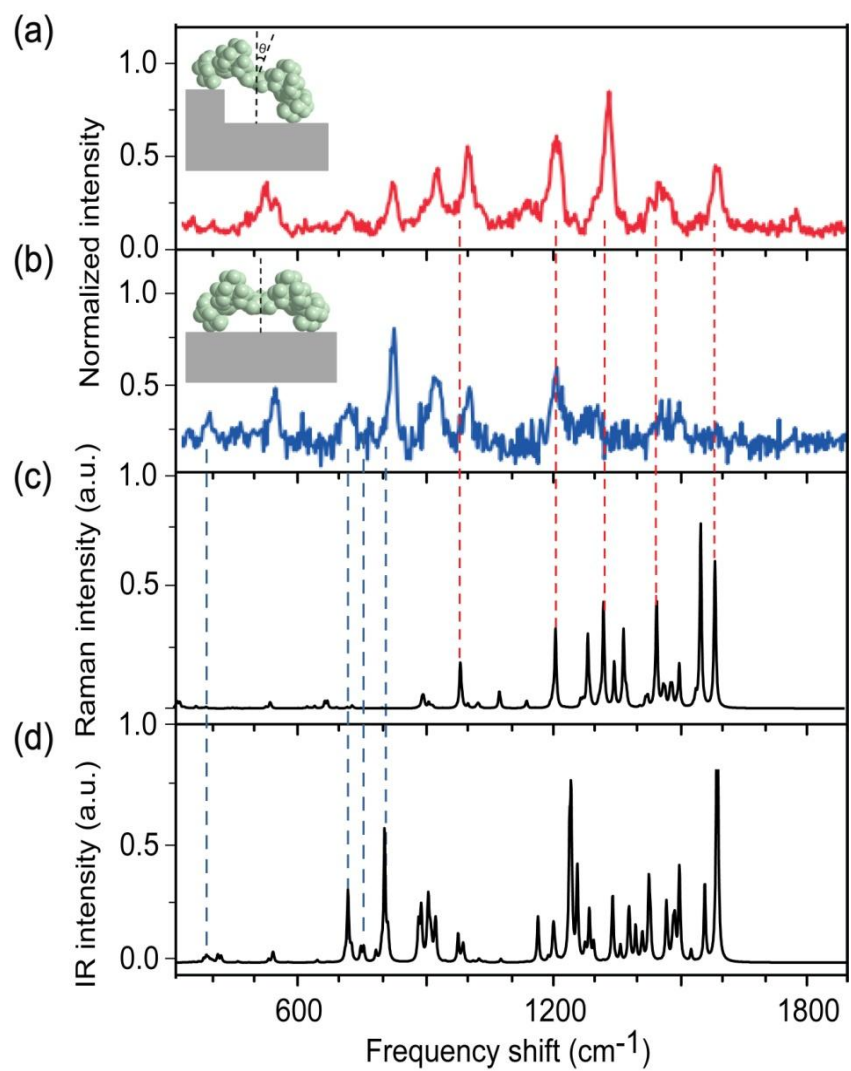

FIG.4. (color online) (a) and (b) Single-molecule TERS spectra for an isolated H$_2$TBPP molecule adsorbed on the terrace or at the step edge. (c) and (d) Calculated TERS spectra of H$_2$TBPP. The inset image in (a) shows the schematic image of tilted H$_2$TBPP molecule with tilt angle of 30°. The inset image in (b) shows the schematic image of a flat-lying H2TBPP molecule. (Adapted from Ref. 1 with permission of American Nature, Copyright 2013).

We provide more evidence of the important contribution of electric field gradient by showing the existence of infrared active modes of molecules in the TERS spectroscopy. Figure 4 shows the experimental TERS spectra of H$_2$TBPP molecules from Ref.1 together with simulated Raman- and IR- active modes. It can be seen clearly that there are four IR-active modes at the low frequency shift (see blue dotted vertical lines) and five Raman-active modes at the high frequency shift (see red dotted vertical lines). As the IR-active modes are only attributed from the electric field gradient effect, these results provide a strong evidence of its important contribution to the Raman signals. Furthermore, the intensity ratio of the IR-active mode to the Raman active mode for flat-lying $H_2TBPP$ molecules is stronger than that for molecules perpendicularly adsorbed on the substrate. That agrees well with the variation trend in Fig. 4 (a) and (b) where as the tilt angle of the molecule varies from $0°$ to $30°$, the relative intensity for IR-active modes decrease while for the Raman-active modes increases. This result demonstrates that the electric field gradient plays a dominant role for a horizontal orientation of molecules; while the electric field is more important in the vertical orientation.

In our previous theoretical calculations, it has been revealed that the spatial resolutions from the contribution of horizontal electric field and electric field gradient are 1 nm for the flat-lying $H_2TBPP$ molecules. If considering the contribution of vertical electric field and electric field gradient, the spatial resolutions are within 2 nm (see Fig. S2). However, as shown in Fig. S1(f), the maximum ratio of $I_2/I_1$ is only 0.14 which reveals that the electric field but not the electric field gradient has a dominant contribution to the ultra-high spatial resolution of TERS in the vertical orientation. Note that the tip diameter in our calculations is set at 2 nm. If we further decrease the tip size, the spatial resolution should be higher. In fact, the size of tip apex in a real STM based TERS system is only several atoms.

In this letter, we propose a reasonable physical mechanism of ultra-high spatial resolution of TERS based on electric field gradient effect. The cone angle of the tip has a significant influence on the electric field enhancement, electric field gradient enhancement and spatial resolution. For the flat-lying $H_2TBPP$ molecules, the electric field gradient has a dominant contribution to the TERS resolution yielding an ultra-high spatial resolution as high as 1 nm. Our work opens up new realms for investigation of ultra-spatial resolution TERS.

This work was supported by the National Natural Science Foundation of China (91436102, 11374353, 11474141, 11474239 and 21173171) and the Program of Liaoning Key Laboratory of Semiconductor Light Emitting and Photocatalytic Materials.

* mtsun@iphy.ac.cn (M. T. Sun), jnchen@iphy.ac.cn (J. N. Chen) or zlyang@xmu.edu.cn (Z. L. Yang).

# Supporting Information

## Effects of Electric Field Gradient on Sub-nanometer Spatial Resolution of Tip-enhanced Raman Spectroscopy

Lingyan Meng,[1,2] Zhilin Yang,[2,*] Jianing Chen,[1,3,*] Mengtao Sun[1,*]

[1]*Beijing National Laboratory for Condensed Matter Physics, Institute of Physics, Chinese Academy of Sciences, Beijing, 100190, China*

[2]*Department of Physics, Xiamen University, Xiamen, 361005, China*

[3] *Collaborative Innovation Center of Quantum Matter, Beijing, 100190, China*

PACS numbers: 73.20.Mf, 78.30.Fs, 87.57.cf, 14.20.Gk

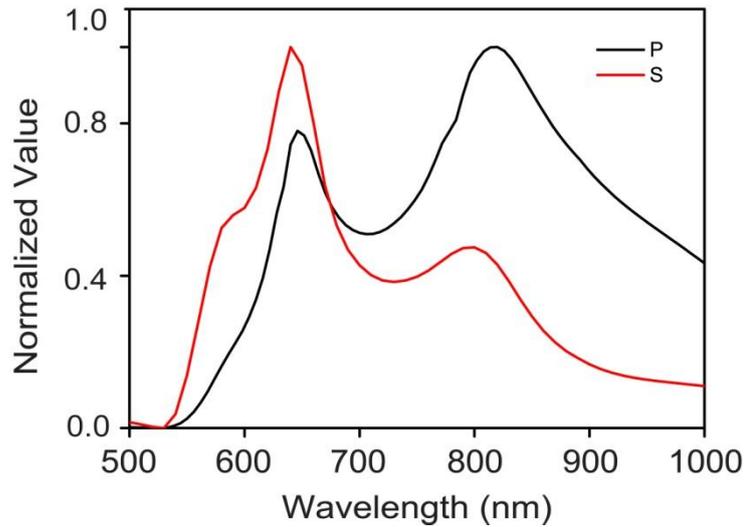

FIG.S1 (color online) Normalized total electric field enhancement (|M|) as a function of wavelength and polarization orientation, where d=1 nm, $\phi=20°$.

FIG.S2 shows the dependence of electric field enhancement on the wavelength and the polarization orientation. Note that the intensities were normalized. For the P-polarization, the relative intensity of resonance peak at longer wavelength is higher than that at the shorter wavelength due to the strong coupling effect between the tip and substrate. While we change the polarization orientation to S-polarization, the resonance peak at longer wavelength is suppressed leading to a lower relative intensity than that of resonance peak at shorter wavelength. Therefore, two resonance peaks are attributed to the vertical (long wavelength) and the horizontal (short wavelength) dipole resonance, respectively.

When the molecules perpendicular adsorbed on the substrate, the vertical electric field and its gradient along the tip axis play the most important role. In Fig.S1 we study the vertical electric field and its gradient field distributions, and their contributions to the TERS resolution in the horizontal orientation at the wavelength 632.8 nm. Fig. S1(a) shows the highest electric field enhancement ($|M_z|$) occurs at the center of the 2D plane showing the maximum field enhancement of 636. As shown in Fig. S1(b) the strongest electric field gradient ($|\nabla M_z|$) is distributed within the ring band 0.7<r<1.7 nm which is the same as that in the horizontal orientation. This can be inferred by the Laplace's equation,

$$\frac{\partial^2 E}{\partial x^2} + \frac{\partial^2 E}{\partial y^2} + \frac{\partial^2 E}{\partial z^2} = 0 \tag{5}$$

Fig. S1(c) shows the distribution of the ratio of electric field gradient enhancement over electric field enhancement ($|\nabla M_z / M_z|$) where the maximum ratio is only 0.036. Fig. S1(d)-(f) shows profiles of $I_1$, $I_2$ and $I_1/I_2$ plotted as function of the lateral displacement of the midway on the 2D plane. As shown in Fig.S1(d), the maximum $I_1$ is $1.6 \times 10^{11}$ and the spatial resolution is 1.5 nm. Fig. S1(e) shows that the maximum $I_2$ is $1.2 \times 10^{10}$ and the spatial resolution 2 nm. In Fig. S1(f), the maximum ratio of $I_2/I_1 \approx 0.14$ which reveals a minor contribution of electric field gradient to the ultra-high spatial resolution of TERS.

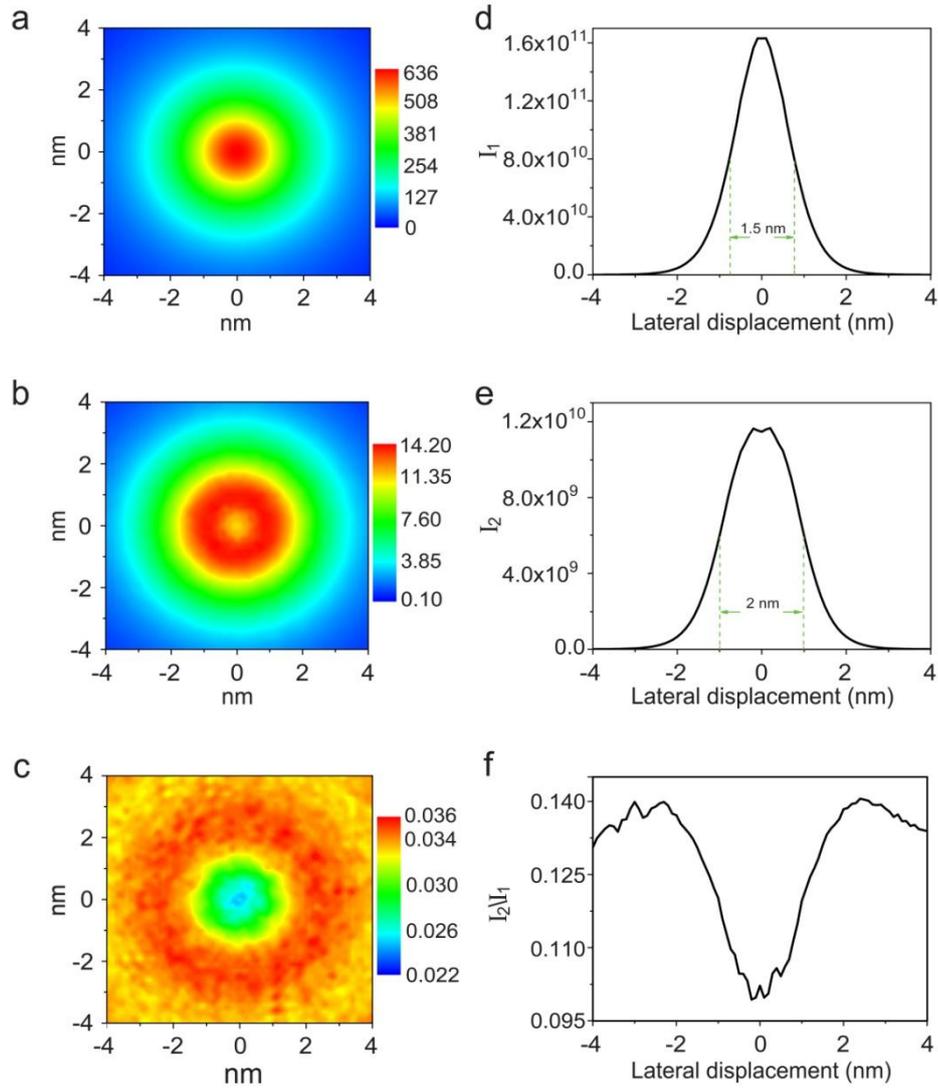

FIG.S2 (color online) (a) The vertical electric field and (b) vertical electric field gradient distribution of the plane between the tip and substrate, where d=1 nm, $\phi$=20°. (c) The ratio of (b) over (a). (d) and (e) The $I_1$, $I_2$ and $I_2/I_1$ is plotted as a function of the lateral displacement which can reveal the spatial resolution of $I_1$, $I_2$ by full width at half maximum (FWHM). (f) $I_2/I_1$ is plotted as a function of the lateral displacement which describes the ratio of electric field to its gradient. The electric gradient is in unit of au which is atomic unit. The excitation wavelength is 632.8 nm.